\documentclass{aa}
\usepackage{graphicx}
\usepackage{times,verbatim}
\usepackage[fleqn]{amsmath}
\begin{document}
\title{The jet-disk symbiosis without \textit{maximal jets}: 
\\1-D hydrodynamical jets revisited}
\titlerunning{1-D Hydrodynamic Jets}
\author{Patrick Crumley\inst{1}, Chiara Ceccobello\inst{1}, Riley M. T. Connors\inst{1}, Yuri Cavecchi\inst{1,2,3}}
\authorrunning{P. Crumley, C. Ceccobello, R. M. T. Connors, Y. Cavecchi}

   \institute{$^1$Anton Pannekoek Institute for Astronomy, University of Amsterdam, P.O. Box 94249, 1090 GE Amsterdam, the Netherlands\\
   $^2$Department of Astrophysical Sciences, Princeton University, Peyton Hall, Princeton, NJ 08544, USA\\
$^3$Mathematical Sciences and STAG Research Centre, University of Southampton, SO17 1BJ, UK\\
              \email{p.k.crumley@uva.nl}
             }
\date{\today}

\abstract{In this work we discuss the recent criticism by \citet{2016A&A...586A..18Z} of the \textit{maximal jet} model derived in \citet{1995A&A...293..665F}. We agree with Zdziarski that in general a jet's internal energy is not bounded by its rest-mass energy density. We describe the effects of the mistake on conclusions that have been made using the \textit{maximal jet} model and show when a \textit{maximal jet} is an appropriate assumption. The \textit{maximal jet} model was used to derive a 1-D hydrodynamical model of jets in \texttt{agnjet}, a model that does multiwavelength fitting of quiescent/hard state X-ray binaries and low-luminosity active galactic nuclei.  We correct algebraic mistakes made in the derivation of the 1-D Euler equation and relax the \textit{maximal jet} assumption. We show that the corrections cause minor differences as long as the jet has a small opening angle and a small terminal Lorentz factor. We find that the major conclusion from the \textit{maximal jet} model, the jet-disk symbiosis, can be generally applied to astrophysical jets. We also show that isothermal jets are required to match the flat radio spectra seen in low-luminosity X-ray binaries and active galactic nuclei, in agreement with other works.}
\maketitle
\section{Introduction}
In this work we reexamine the \textit{maximal jet} model of accreting black-holes with disks and jets derived in \citet{1995A&A...293..665F}. The idea behind a \textit{maximal jet} is that there is a strict upper limit on the power carried by a jet in terms of its mass flux. As first pointed out by \citet{2016A&A...586A..18Z}, the \textit{maximal jet} is the result of an erroneous conclusion that the internal energy density of a gas must be less than or equal to the rest-mass energy density of the gas. The \textit{maximal jet} model was used to link the power carried by the jet to the power in the accretion disk, arriving at the highly influential and widely used ``jet-disk symbiosis.'' The main result of the jet-disk symbiosis states that the total power in the jet, $L_j$, can be related to the mass accretion rate of the disk, $\dot{M}_{\rm disk}$, through the jet's Lorentz factor $\gamma_j$ and an efficiency, $\eta<1$. $L_j = \eta \gamma_j \dot{M}_{\rm disk} c^2$. 

Without appealing to a \textit{maximal jet}, we argue in this paper that the jet-disk symbiosis is reasonable for astrophysical accreting black-holes in general. In fact, it is reasonable to estimate the power of the jet to be $\lesssim \eta \dot{M}_{\rm disk}c^2$, as long as one takes $\eta \lesssim {\rm a\ few}$, as opposed to being strictly less than one. It is intuitive why this conclusion should hold in a Blandford-Payne type jet where the disk itself is powering the jet \citep{1982MNRAS.199..883B}, but the conclusion should also hold for jets that are powered by the black hole's rotational energy via the Blandford-Znajek mechanism \citep{1977MNRAS.179..433B}. In the Blandford-Znajek mechanism, the jet power is proportional to the poloidal magnetic flux at the black hole, and the magnetic flux that can be carried to the black hole is ultimately limited by the mass-accretion rate \citep{2003PASJ...55L..69N, 2011MNRAS.418L..79T}. This explains why the jet-disk symbiosis has been such a successful concept.

However, the \textit{maximal jet} conclusion results solely from an algebraic mistake in \citet{1995A&A...293..665F} and cannot be applied broadly to accreting black-holes with jets. We argue in this work that if the jet is efficiently accelerated and has a small terminal Lorentz factor, the initial enthalpy should be roughly equal to the rest-mass energy, in agreement with a \textit{maximal jet}. A jet with a large terminal Lorentz factor will start with a large enthalpy, but if the jet is efficiently accelerated, the internal energy of the jet will be approximately equal to the rest-mass energy density after the jet has reached its final Lorentz factor \citep{2003ApJ...596.1080V, 2006MNRAS.368.1561M}. { There are some cases when this approximation breaks down. For a steady-state, axisymmetric, magnetically-accelerated outflow to be efficiently accelerated, it must stay causally connected in the transverse direction. However, it is unclear if this requirement is an actual impediment to the magnetic acceleration of astrophysical jets, or an artifact of the symmetries imposed. (For a concise review of magnetic acceleration of jets, see \citeauthor{2011MmSAI..82...95K} \citeyear{2011MmSAI..82...95K}.) A radiation or thermal pressure-driven jet will be efficiently accelerated even if it is conical and free-streaming. We assume the jet opening angle is small enough to ensure it remains in causal contact with the external medium throughout this paper.}

In addition, we correct algebraic errors made in the derivation of the jet's Lorentz factor as a function of distance in \citet{1996ApJ...464L..67F}. This Lorentz factor profile is used to calculate the dynamics of a jet in \texttt{agnjet}, an outflow dominated model of low-luminosity accreting black holes that has been applied to several different low-luminosity active galactic nuclei and X-ray binaries \citep{2005ApJ...635.1203M, 2008ApJ...681..905M, 2009MNRAS.398.1638M, 2010MNRAS.409..763V, 2015ApJ...812L..25M, 2015MNRAS.446.4098P, 2016arXiv161200953C}. The physics behind \texttt{agnjet} is presented in detail in \citet{2005ApJ...635.1203M} \& \citet{2009MNRAS.398.1638M}. We characterize how the changes to the Lorentz factor profile effect the resulting spectral energy distribution calculated by \texttt{agnjet}. { We find that the aforementioned algebraic mistakes have a negligible effect on the radiation from the outflow as long as the jet is roughly isothermal, has a small Lorentz factor, and is launched with an aspect ratio of order unity.}

Our paper is organized as follows: in Section \ref{BernEq} we outline the mistake made when the \textit{maximal jet} was derived. Then, we argue why the main conclusion of the \textit{maximal jet} model, the jet-disk symbiosis, still holds. We also describe the systems for which the \textit{maximal jet} model can be applied. In Section \ref{Jets}, we re-examine 1-D pressure-driven jets, relaxing the \textit{maximal jet} requirement. In Section \ref{AGNJET}, we make the dynamics of \texttt{agnjet} self-consistent, and find the combined effects of all the changes to the dynamics on the calculated spectral energy distribution are small. We end by summarizing and discussing our results.

\section{Bernoulli's equation and Maximal Jets}
\label{BernEq}
The total power of an axisymmetric, conical jet at height $z$ from the launching point with an opening angle $\theta$ is equal to the jet's Lorentz factor $\gamma_j$ times the enthalpy flux,
\begin{equation}
\label{eq:Luminosity}
L_j=\gamma_j^2\beta_j c \omega \pi z^2\sin^2{\theta},
\end{equation}
where $\omega$ is the enthalpy. In ideal magneto-hydrodynamics (MHD), a baryonic jet with a co-moving number density of protons $n$, has an enthalpy given by
\begin{equation}
\omega = n m_p c^2 + U_j + P_j =  n m_p c^2 + U_{\rm th} + P_{\rm th} + \frac{B^2}{4\pi}.
\end{equation}
$U_j$, $P_j$ are the total energy density and pressure of the jet, which can be broken down into a gas component ($U_{\rm th}$, $P_{\rm th}$) and a magnetic component ($U_{B} =  P_{B} = B^2/8\pi$).\footnote{The gravitational potential energy and the radiation pressure and energy density contributions to the enthalpy are neglected in this work.} The jet's gas pressure can be related to the internal energy of the gas via the adiabatic index, $\Gamma$, $P_{\rm th} = (\Gamma-1)U_{\rm th}$. We define the magnetization parameter, $\sigma$, as $B^2/(4\pi n m_p c^2)$.\footnote{Our $\sigma$ is the same as the more standard magnetization parameter $\sigma$ in the ``force-free'' MHD regime. We neglect the gas pressure contributions to $\sigma$ to simplify the equations.} The enthalpy then simplifies to
\begin{equation}
\label{eq:Enthalpy}
\omega = n m_p c^2 \left[1+\sigma + \frac{\Gamma U_{\rm th}}{n m_p c^2}\right]
\end{equation}
In \citet{1995A&A...293..665F}, the authors assume that the magnetic fields are isotropically turbulent, and therefore can be treated as an ideal gas with adiabatic index $\Gamma$. The authors then write the enthalpy in terms of the total jet internal energy density $U_j$ as
\begin{equation}
\omega \sim n m_p c^2 + \Gamma U_j
\label{eq:EnthalpyNoSigma}
\end{equation}
The authors then re-write the enthalpy in terms of the sound speed, $\beta_s$, adiabatic index, and density. The sound speed is 
\begin{equation}\label{eq:SoundSpeed}
\beta_s^2 = \frac{\Gamma P_j}{\omega} = \frac{\Gamma(\Gamma-1)U_j}{\omega},
\end{equation}
\citep[see e.g.][]{1980PhFl...23.1083K}.
From Equations (\ref{eq:EnthalpyNoSigma}) \& (\ref{eq:SoundSpeed}), one can derive the well known result that the maximal sound speed is $\sqrt{\Gamma-1}$. Substituting Equation (\ref{eq:SoundSpeed}) into Equation (\ref{eq:EnthalpyNoSigma}) and solving for $\omega$ yields
\begin{equation}\label{eq:omegaSound}
\omega = \frac{n m_p c^2}{1-\beta_s^2/(\Gamma-1)}
\end{equation}
Compare the above equation to the equivalent equation in \citet{1995A&A...293..665F}. They give a formula that is the approximation when $\beta_s$ is small,
\begin{equation}\label{eq:omegaNonRelSound}
\omega \approx n m_p c^2\left(1+\frac{\beta_s^2}{\Gamma-1}\right) \quad {\rm when} \quad \beta_s \ll \sqrt{\Gamma-1}.
\end{equation}
The mistake is that \citet{1995A&A...293..665F} then use Equation (\ref{eq:omegaNonRelSound}) to argue $\omega$ must be less than $2n m_p c^2$ because $\beta_s$ must be less than $\sqrt{\Gamma-1}$. Clearly this is wrong, because when using the correct formula, Equation (\ref{eq:omegaSound}),  $\omega$ diverges to infinity as $\beta_s \rightarrow \sqrt{\Gamma-1}$.
This mistake was first pointed out by \citet{2016A&A...586A..18Z}.

Since the total jet power is equal to the Lorentz factor times the enthalpy flux, the mistake leads to a maximal jet power $L_{j} \leq 2\gamma_j \dot{M}_{j}c^2$, where $\dot{M}_{j}$ is the mass flux through the jet. Protons are not created at the jet base, so \citet{1995A&A...293..665F} argue $\dot{M}_j\leq\dot{M}_{\rm disk}$, where $\dot{M}_{\rm disk}$ is the mass accretion rate of the disk. Therefore, \citet{1995A&A...293..665F} conclude  $L_j \sim \eta \gamma_j \dot{M}_{\rm disk} c^2$, $\eta \leq 1$, i.e., the jet-disk symbiosis. The mass flux through the jet could be larger than $\dot{M}_{\rm disk}$ if the jet has significant baryon loading from winds. If the disk drives a strong wind, the disk wind may be able to increase the baryon loading, but it is generally thought that in X-ray binaries a strong disk wind is associated with the soft states, i.e., when the jet is not present \citep{2009Natur.458..481N}. New evidence shows simultaneous winds and jets in the high state, but there is no evidence that shows a strong wind during the low hard-state \citep{2016Natur.534...75M,2016MNRAS.463..615K,2016arXiv161008517M} Stellar winds could also play a role in increasing $\dot{M}_j$ in AGN or high-mass X-ray binaries \citep{1994MNRAS.269..394K}. At any rate, the baryon loading would only significantly increase the power carried by the jet when the power of the intercepted baryons exceeds that of the jet.  

It is a reasonable approximation that the power of the jet does not significantly exceed $\dot{M}_{BH}c^2$ even if the power in the jet is supplied by the spin of the black hole via the Blandford-Znajek mechanism. In the Blandford-Znajek mechanism, the power extracted is proportional to the magnetic flux carried to the black hole \citep{1977MNRAS.179..433B}. The magnetic flux carried to the black-hole is proportional to $\dot{M}_{BH}$, and GRMHD simulations find that the resultant jet power exceeds $\dot{M}_{BH}c^2$ by a factor of at most $\sim 3$ \citep{2011MNRAS.418L..79T}. However, there is no general requirement that the resultant jet's power is dominated by the jet's kinetic energy throughout the jet. For instance, in a Poynting-dominated jet, the power is $L_j = \gamma_j\dot{M}_{j}c^2(1+\sigma)$ and $\sigma$ can be $\gg 1$. Clearly, the enthalpy flux can be much larger than the mass flux, but in the next subsection, we will argue from completely different grounds than \citet{1995A&A...293..665F} that while a jet can start with any initial enthalpy, the jet will reach an $\omega/n m_p c^2 \sim 2$ at the modified magneto-sonic fast point if it is efficiently accelerated. Furthermore, in efficiently accelerated jets that have mildly relativistic terminal Lorentz factors, i.e., $\gamma_j\beta_j\sim 1$, the initial enthalpy will not exceed the rest mass energy density by a significant amount. The reason why this is true can be seen most easily via Bernoulli's equation. 

\subsection{Bernoulli's equation}
In a steady-state, conservative jet, the total power carried by the jet is constant along the jet. Dividing the power by another conserved quantity, the particle number flux through a cross-section of the jet, one re-derives the relativistic Bernoulli's equation \citep[see e.g.][]{1980PhFl...23.1083K}:
\begin{equation}
\label{eq:Bernoulli}
\gamma_j \frac{\omega}{n} = {\rm constant}.
\end{equation}
Bernoulli's equation is simply a statement that the energy of the fluid per particle must not change as the particles travel along a streamline, as long as particles are not created or destroyed. In this paper we are only considering self-similar jets with no gradient of the pressure in the toroidal direction of the jet. If a jet is launched with an initial Lorentz factor $\gamma_0$ with an initial enthalpy per particle of $\omega_0/n_0$, then because $\omega/n \geq m_p c^2$, it is clear from Eq (\ref{eq:Bernoulli}) and (\ref{eq:Enthalpy}) there is a maximal Lorentz factor:
\begin{equation}
\label{eq:MaxLF}
\gamma_{\rm max}  = \gamma_0\frac{\omega_0}{n_0m_p c^2} = \gamma_0 \left[1+\sigma_0 + \frac{\Gamma U_{\rm th,0}}{n_0 m_p c^2}\right].
\end{equation}
If a jet achieves $\gamma_{\rm max}$, it means it has converted 100\% of its Poynting and thermal energy into bulk kinetic energy. { Doing so in a steady-state, Poynting-dominated jet requires that the jet be causally connected in the transverse direction.}\footnote{ The requirement that Poynting-dominated jets remain causally connected in the transverse direction can be relaxed if the outflow is impulsive \citep{2011MNRAS.411.1323G}}
{ If the jet goes out of causal contact in the transverse direction, the  magnetic pressure may not be able to accelerate the jet further.} If the flow is completely spherical with $\sigma_0\gg1$, the outflow reaches a terminal Lorentz factor $\sim \sigma_0^{1/3}$ \citep{1970ApJ...160..971G}. If instead the jet has an opening angle $\theta_j$, the terminal Lorentz factor is $\gamma_f = \min{\left(\sigma_0,\sigma_0^{1/3}\theta_j^{-2/3}\right)}$ \citep{2015PhR...561....1K}. { When the jet is thermally or radiatively driven, equation (\ref{eq:MaxLF}) should hold regardless of the jet's geometry.}
\citet{2003ApJ...596.1080V, 2007MNRAS.380...51K} show that in some Poynting dominated jets, electromagnetic fields will naturally self-collimate the jet and ensure that the jet remains causally connected up to the modified magneto-sonic fast point. In doing so the jet reaches rough equipartition between kinetic flux and Poynting flux, and the jet reaches a final Lorentz factor equal to $\sim \frac{1}{2}\omega_0/n_0 m_p c^2$.\footnote{The jet is launched with a sub-relativistic velocity, $\gamma_0 = 1$}
Therefore, using Bernoulli's equation, the jet should have $\omega/n m_p c^2\sim 2$ at the modified magneto-sonic fast point, right after the jet has finished accelerating. Furthermore, $\omega_0/n_0 m_p c^2 \approx 2$ is likely to be a good assumption for mildly-relativistic outflows with $\gamma_j\beta_j$ of order unity, because if $\omega_0\gg n_0 m_p c^2$  the jets would reach $\gamma_j$ that are too large. Mildly relativistic jets are expected to occur in quiescent/hard state black-hole x-ray binaries\footnote{Although see \citet{2004MNRAS.355L...1H,2006MNRAS.367.1432M} regarding the difficulties in placing a strong upper limit on the Lorentz factor of X-ray binaries.} \citep{2003MNRAS.344...60G, 2007MNRAS.375.1087M}. A mildly relativistic jet may be launched by the super-massive black-hole at the center of our galaxy, Sgr A* \citep{2009A&A...496...77F,2015A&A...576A..41B}.

Therefore, $\omega_0/n_0 \sim 2m_p c^2$ should be a fine assumption as long as we restrict ourselves to mildly relativistic outflows with small enough opening angles. In jets with large Lorentz factors, $\gamma_j\beta_j\gg 5$, like blazars, BL Lacs, relativistic tidal disruption events, and gamma-ray bursts, $\omega_0/n_0 \gg 2 m_p c^2$, but if the jet is accelerated efficiently, at the modified magneto-sonic fast point $\omega/n m_p c^2 \sim 2$.\footnote{If the highly relativistic jet is comprised of electron-positron pairs, then $\omega_0/n_0 \gg 2 m_e c^2$.} The previous jet model of \texttt{agnjet}, and the one described in this paper, are not capable of reproducing the jets in these objects.

In the process of examining the \textit{maximal jet} model, we have discovered additional minor algebraic mistakes in the Lorentz factor profile derived in \citet{1996ApJ...464L..67F} and used in \texttt{agnjet}. In the rest of the paper, we show that these mistakes are minor and do not affect the conclusions drawn from fitting the model to low Lorentz factor sources. We then show that isothermal or nearly isothermal jets are required to have a flat radio spectrum.

\section{Pressure-Driven Conical Jets}
\label{Jets}
\begin{figure}
\includegraphics[width = 0.5\textwidth]{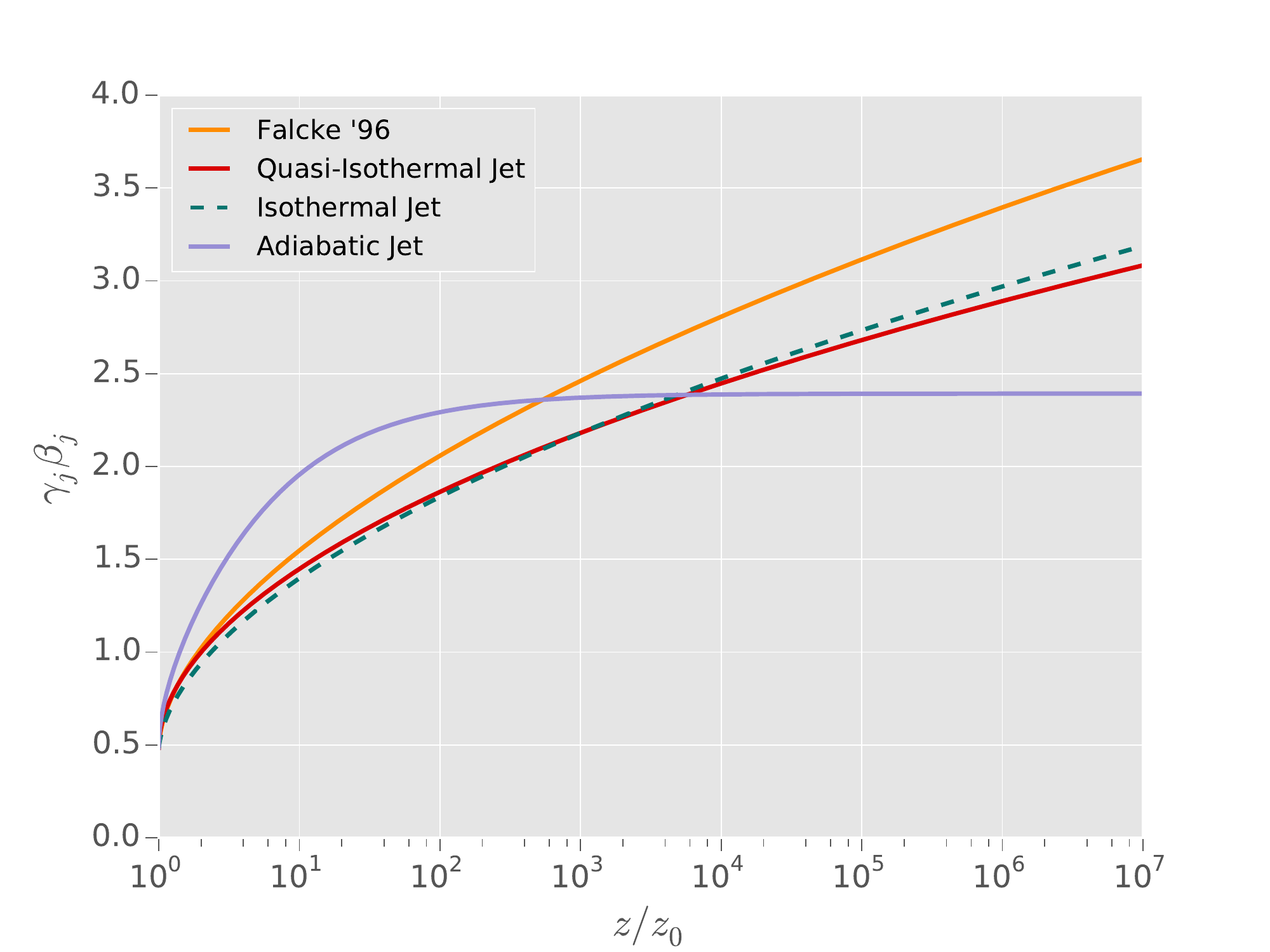}
\caption{This figure shows the difference between the Lorentz factor profile derived in \citet{1996ApJ...464L..67F} and used in \texttt{agnjet} c.f. \citet{2005ApJ...635.1203M} (yellow solid line), the derived Lorentz factor profile after correcting for algebraic mistakes  (eq \ref{eq:AGNJET_Corrected}, red solid line), the 1-D Euler equation in a conical jet assuming continual heating of the jet by an outside source such that the jet is isothermal, $U_j = n m_p c^2$ (green dashed line), and the Euler equation in an adiabatic jet, where $U_j = n_0 m_p c^2 (n/n_0)^{\Gamma}$ (lilac solid line). If the jet follows Bernoulli's equation, $\gamma_{\rm max}\beta_{\rm max} \approx 2.39$ when $\gamma_0\beta_0 \approx 0.485$ and $\Gamma = 4/3$---precisely the terminal value in the adiabatic jet Lorentz factor profile.}
\label{fig:AGNJETvEuler}
\end{figure}

In this section we reproduce with minor corrections the derivation of the one-dimensional propagation of a quasi-isothermal hydrodynamic jet from \citet{1996ApJ...464L..67F}.  We find that the corrected quasi-isothermal acceleration profile agrees within $\sim$20\% with the result in \citet{1996ApJ...464L..67F}, and the corrected Lorentz factor profile is much closer to the profile in a perfectly isothermal flow (see Figure \ref{fig:AGNJETvEuler}).

The one-dimensional propagation of a supersonic jet in the $z$ direction follows the Euler equation given in \citet{1973erh..book.....P, 2009A&A...496...77F}:
\begin{equation}
\label{eq:1dEuler}
\gamma_j\beta_j n \frac{\partial}{\partial z}\left\{\gamma_j \beta_j \frac{\omega}{n}\right\} = -\frac{\partial P_j}{\partial z}
\end{equation}
Following \citet{1995A&A...293..665F}, we assume that the jet is well-described by a fluid with adiabatic index 4/3 and use the enthalpy from Eq (\ref{eq:EnthalpyNoSigma})
\begin{equation}
\omega = n m_p c^2 + \Gamma U_j,
\end{equation}
so Eq (\ref{eq:1dEuler}) becomes:
\begin{equation}
\label{eq:Euler_expanded}
\gamma_j\beta_j n \frac{\partial}{\partial z}\left\{\gamma_j\beta_j\left(m_p c^2 +\frac{\Gamma U_j}{n}\right)\right\} = -(\Gamma-1)\frac {\partial U_j}{\partial z}.
\end{equation}

Particle number conservation along the jet forces the number density to be
\begin{equation}
\label{eq:Density}
n_j = n_0 \left(\frac{\gamma_j\beta_j}{\gamma_0\beta_0}\right)^{-1}\left(\frac{z}{z_0}\right)^{-2}\left(\frac{\sin{\theta}}{\sin{\theta_0}}\right)^{-2},
\end{equation}
where $\theta$ is the opening angle of the jet (i.e. the cross-sectional radius of the jet divided by the height of the jet). 

The jet is launched at an initial height of $z_0$, and is assumed to be traveling at the sound speed. 
The sound speed in the jet's rest frame, $\beta_s$, is
\begin{equation}
\beta_s^2 = \frac{\Gamma P_j}{\omega} = \frac{\Gamma(\Gamma-1)U_j}{n m_p c^2 + \Gamma U_j}.
\end{equation}
Instead of requiring $U_j = n m_p c^2$ throughout the jet, we introduce a new parameter $\zeta$, which is the ratio between the initial internal energy of the jet and the initial rest-mass energy density, i.e., $U_{0} = \zeta n_0 m_p c^2$. $\zeta = 1$ corresponds to the maximal jet conditions in \citet{1996ApJ...464L..67F}. The initial sound speed at the base of the jet is
\begin{equation}
\beta_{s0} = \sqrt{\frac{\zeta\Gamma(\Gamma-1)}{1+\zeta\Gamma}}.
\end{equation}
For $\zeta = 1,\ \Gamma = 4/3$ the sound speed is $\approx 0.43$. Since the jet is assumed to be launched traveling at the sound speed, the initial velocity of the jet is 
\begin{equation}
\label{eq:InitialGamBeta}
\gamma_0\beta_0 = 1/\sqrt{\beta_{s0}^{-2}-1} = \sqrt{\frac{\zeta\Gamma(\Gamma-1)}{1+2\zeta\Gamma-\zeta\Gamma^2}}.
\end{equation}
For $\Gamma = 4/3$, $\zeta = 1$, $\gamma_0\beta_0 \approx 0.485$.

The jet Lorentz factor profile from \citet{1996ApJ...464L..67F} can be derived by treating the jet as a conical jet ($\theta = \theta_0$) and using particle number conservation to get a $z$ dependence on the density (see Eq \ref{eq:Density}). To fix the $z$ dependence on the internal energy, we need a prescription of how the temperature changes with a change in density. If the jet is isothermal, $T_j$ is constant and $U_e\propto n$. If the jet is adiabatic, $U_e \sim nkT_j  \propto n^\Gamma$, and therefore $T_j \propto n^{\Gamma-1} \propto (\gamma_j\beta_j)^{1-\Gamma}z^{2-2\Gamma}$. \citet{1996ApJ...464L..67F} assumes that the gas is only able to do $P dV$ work in the $z$-direction, hence the only adiabatic losses are due to the jet's acceleration. We call this assumption quasi-isothermal. If the jet is quasi-isothermal, the temperature $T_j$, is proportional to $(\gamma_j\beta_j)^{1-\Gamma}$. It is difficult to understand how exactly the gas is prevented from doing $P dV$ work in the lateral direction. It is far more realistic to assume there is continuous particle acceleration to counteract the adiabatic losses due to the expansion, as \citet{1979ApJ...232...34B} use to explain their isothermal jet model. For a heating mechanism to recover the quasi-isothermal temperature dependence, it means it must be capable of compensating the large adiabatic losses due to the lateral expansion, but not the comparatively small adiabatic losses from the acceleration. Why the heating mechanism would do this is unclear. We retain $T_j\propto(\gamma_j\beta_j)^{1-\Gamma}$ here for historical reasons, and we note that when using the correct Euler equation, the difference between the quasi-isothermal case and the isothermal case is negligible for the small Lorentz factors achieved in jets with $U_{j,0} \sim n_0 m_p c^2$ and $\Gamma=4/3$, as assumed in this work.

In a quasi-isothermal jet, $U_j$ is
\begin{equation}
\label{eq:U_j_quasi}
U_j = \zeta n_0 m_p c^2\left(\frac{\gamma_j\beta_j}{\gamma_0\beta_0}\right)^{-\Gamma}\left(\frac{z}{z_0}\right)^{-2}.
\end{equation}

Substituting Eqs (\ref{eq:U_j_quasi}) and (\ref{eq:Density}) into Eq (\ref{eq:1dEuler}), and assuming the jet is launched with an initial $\gamma_0\beta_0$ equal to the sound speed (Eq \ref{eq:InitialGamBeta}), the 1-D Euler equation that results is
\begin{equation}
\label{eq:AGNJET_Corrected}
\left\{\gamma_j\beta_j\frac{\Gamma+\xi}{\Gamma-1}-\Gamma\gamma_j\beta_j-\frac{\Gamma}{\gamma_j\beta_j}\right\}\frac{\partial \gamma_j\beta_j}{\partial z} = \frac{2}{z};
%\qquad\xi = \frac{1}{\zeta}\left(\gamma_j\beta_j\sqrt{\frac{1+2\zeta\Gamma-\zeta\Gamma^2}{\zeta\Gamma(\Gamma-1)}}\right)^{\Gamma-1}.
\end{equation}
\begin{equation}
\xi = \frac{1}{\zeta}\left(\frac{\gamma_j\beta_j}{\gamma_0\beta_0}\right)^{\Gamma-1};
\qquad 
\gamma_0\beta_0=\sqrt{\frac{\zeta\Gamma(\Gamma-1)}{1+2\zeta\Gamma-\zeta\Gamma^2}}.
\end{equation}
The above equation should reduce to the jet Lorentz factor profile used in \citet{1996ApJ...464L..67F,2005ApJ...635.1203M} when $\zeta = 1$. However, it differs from Eq (2) in \citet{1996ApJ...464L..67F}:
\begin{equation}
\label{eq:Heino96}
\left\{\gamma_j\beta_j\frac{\Gamma+\xi}{\Gamma-1}-\frac{\Gamma}{\gamma_j\beta_j}\right\}\frac{\partial \gamma_j\beta_j}{\partial z} = \frac{2}{z};
\end{equation}
\begin{equation}\xi = \left(\gamma_j\beta_j\frac{\Gamma+1}{\Gamma(\Gamma-1)}\right)^{1-\Gamma}
\end{equation}
The difference between our equation and the equation in \citet{1996ApJ...464L..67F} can be accounted for as follows: the $-\Gamma\gamma_j\beta_j$ term in Eq (\ref{eq:AGNJET_Corrected}) results from a neglected $\frac{\partial}{\partial z}(U_j/n)$ term, the difference in the exponent in $\xi$ results from an arithmetic error, and finally the difference in the inside of the parenthesis of $\xi$ terms is from setting $\gamma_0\beta_0 = \beta_{s0}^{2}$ instead of using the proper value given in Eq (\ref{eq:InitialGamBeta}). The difference between the solutions of Eqs (\ref{eq:AGNJET_Corrected}) and (\ref{eq:Heino96}) are small and shown in Figure \ref{fig:AGNJETvEuler}. In Figure \ref{fig:AGNJETvEuler}, we also include solutions to the 1-D Euler equations when the jet is isothermal ($T_j= {\rm constant}$, i.e., Eq \ref{eq:Heino96} with $\xi=1$) and adiabatic ($T_j\propto(\gamma_j\beta_j)^{1-\Gamma}z^{2-2\Gamma}$, see Eq \ref{eq:FullEulerAdiabatic}). 

The above quasi-isothermal and isothermal solutions do not conserve energy, nor do they follow the Bernoulli equation. The violation of Bernoulli's equation is clear by looking at maximal Lorentz factor on any pressure driven jet that conserves energy, with $U_0 = \zeta n_0 m_p c^2$, Eq (\ref{eq:MaxLF}) becomes
\begin{equation}
\label{eq:MaxGamma}
\gamma_{\rm max} = \gamma_0 (1+\Gamma\zeta),
\end{equation}
or $\gamma_{\rm max} = 7\gamma_0/3$ for a relativistic gas starting with equal parts internal energy density and rest mass energy density, $U_{0}=n_0 m_p c^2$. All solutions  except for the adiabatic solution eventually reach a $\gamma_j$ that exceeds $\gamma_{\rm max}$, which is easily seen in Figure \ref{fig:AGNJETvEuler}. 

\begin{figure}
\includegraphics[width = 0.5\textwidth]{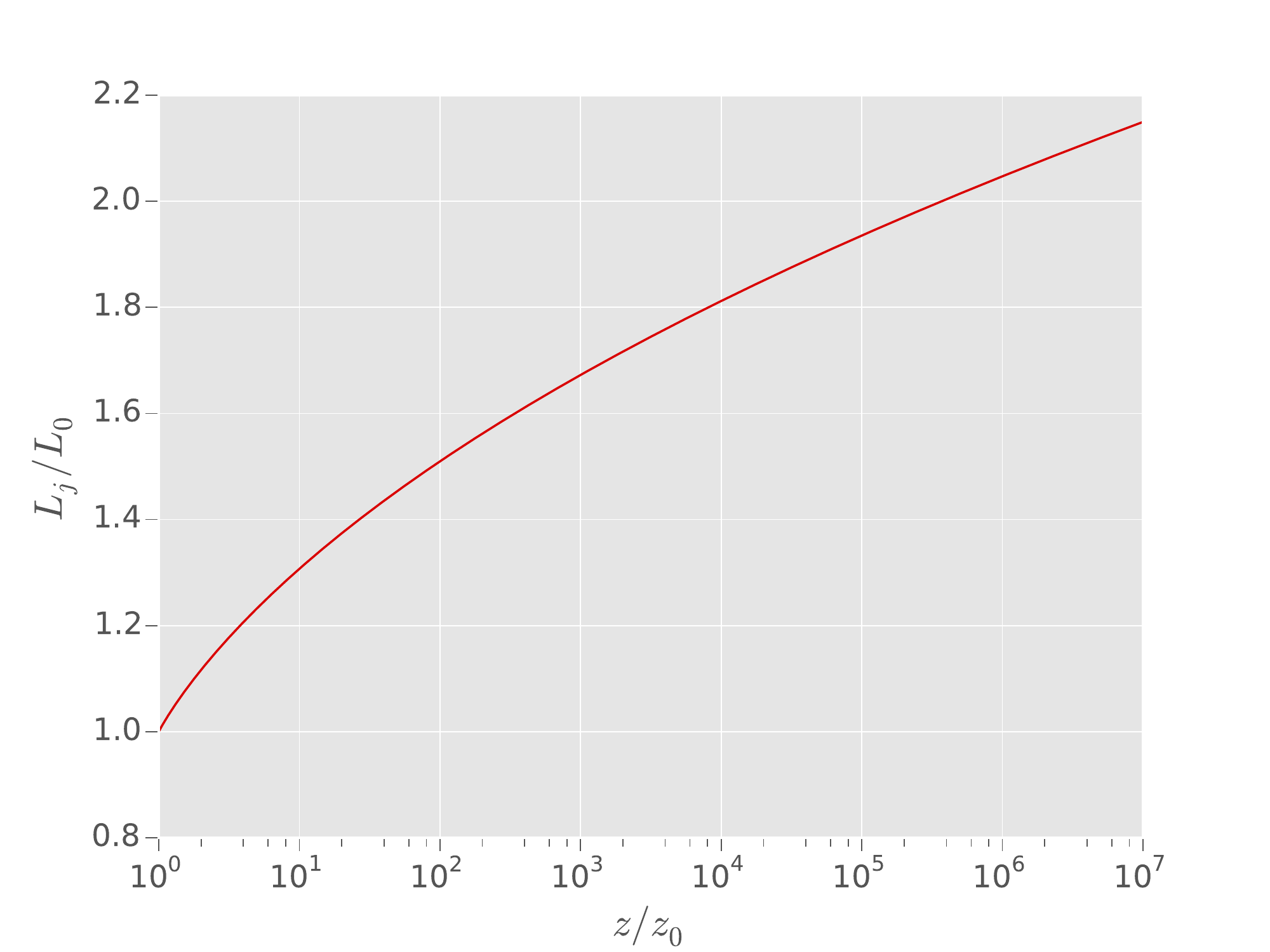}
\caption{This plot shows the total power carried by a quasi-isothermal jet as a function of distance from the black hole $z$ normalized to its initial power. The power of the jet increases as it propagates (see Equation \ref{eq:Heating}). One possibility to explain how the jet is heated is via some internal process not captured by the time-independent Euler equation, such as internal shocks.}
\label{fig:AGNJET_Heating}
\end{figure}

The total amount of heating needed to explain the solution in \texttt{agnjet} is equivalent to the increase in the jet power. Using Eq (\ref{eq:Luminosity}) for the jet's total power, the increase of power in a quasi-isothermal jet is
\begin{equation}
\label{eq:Heating}
\frac{L_j}{L_0} = \frac{1}{1+\Gamma}\frac{\gamma_j}{\gamma_0}\left[1+\Gamma\left(\frac{\gamma_j\beta_j}{\gamma_0\beta_0}\right)^{1-\Gamma}\right].
\end{equation}
The required heating to power a quasi-isothermal jet is shown in Figure \ref{fig:AGNJET_Heating}. The heating must come from some internal process in the jet, but the heating mechanism is not capable of being captured by our time-independent, laminar flow treatment in this work. One possibility is that heating originates from internal shocks \citep{2013MNRAS.429L..20M}. Internal shocks would do more than just heat the gas, they would also change the momentum and hence the dynamics. Magnetic reconnection can convert magnetic energy into thermal energy to keep the jet's electrons isothermal, but reconnection would not increase the total power carried by the jet.

The jet will conserve energy if instead of assuming $T_j\propto (\gamma_j\beta_j)^{1-\Gamma}$, we use an adiabatic jet where $U_j\propto n^\Gamma$, or $T_j\propto (\gamma_j\beta_j)^{1-\Gamma}z^{2-2\Gamma}$. In an adiabatic conical jet the $z$ dependence of the internal energy is:
\begin{equation}
U_j = \zeta n_0 m_p c^2 
\left(\frac{\gamma_j\beta_j}{\gamma_0\beta_0}\right)^{-\Gamma}
\left(\frac{z}{z_0}\right)^{-2\Gamma},
\end{equation}
and the full 1-D Euler equation is
\begin{equation}
\label{eq:FullEulerAdiabatic}
\left\{\gamma_j\beta_j\frac{\Gamma+\xi}{\Gamma-1}-\Gamma \gamma_j\beta_j-\frac{\Gamma}{\gamma_j\beta_j}\right\}\frac{\partial \gamma_j\beta_j}{\partial z} = \frac{2\Gamma}{z}\left(1+\gamma_j^2\beta_j^2\right);\\
\end{equation}
\begin{equation}
\xi = \frac{1}{\zeta}\left(\gamma_j\beta_j\sqrt{\frac{1+2\zeta\Gamma-\zeta\Gamma^2}{\zeta\Gamma(\Gamma-1)}}\right)^{\Gamma-1}\left(\frac{z}{z_0}\right)^{2(\Gamma-1)}.
\end{equation}
The solution to the adiabatic 1-D Euler equation when $\zeta = 1$ is shown in Figure \ref{fig:AGNJETvEuler}. Unlike the quasi-isothermal case, the jet reaches the maximal Lorentz factor equal to that predicted by the Bernoulli equation $\gamma_0(1+\Gamma)$.
%Different solutions to the adiabatic 1-D Euler equation are shown in Figure \ref{fig:Adiabatic}. Unlike the quasi-isothermal case, the jet Lorentz factor profile changes drastically with $\zeta$. In all of the curves, the final Lorentz factor of the jet is equal to $\gamma_0(1+\Gamma\zeta)$. 

\section{Radiation, Collimation, and a more Self-Consistent \texttt{agnjet}}
\label{AGNJET}
The \texttt{agnjet} model was developed in \citet{2005ApJ...635.1203M,2009MNRAS.398.1638M} as way of fitting multi-wavelength spectra of black-hole jets. The model is described in full detail in the aforementioned papers, but we will give a brief description here. In \texttt{agnjet} the gas is assumed to be moving at a velocity equal to the sound speed though a nozzle with constant radius $r_0$ that ends at $z_0$. At $z_0$, the jet is allowed to expand at constant velocity equal to the sound speed\footnote{$\gamma_s\beta_s =\gamma_0\beta_0$} in the cross-section radial direction, and weakly accelerated in the $z$ direction by the jet's pressure, i.e., it follows the 1-D Euler equation in the $z$ direction. Electrons initially have a Maxwell-Juttner thermal distribution in the nozzle and the jet, and a fraction of the electrons are accelerated into a non-thermal population at a height $z_{\rm acc}$. The electrons radiate via synchrotron and inverse Compton processes. 

\begin{figure}
\includegraphics[width = 0.5\textwidth]{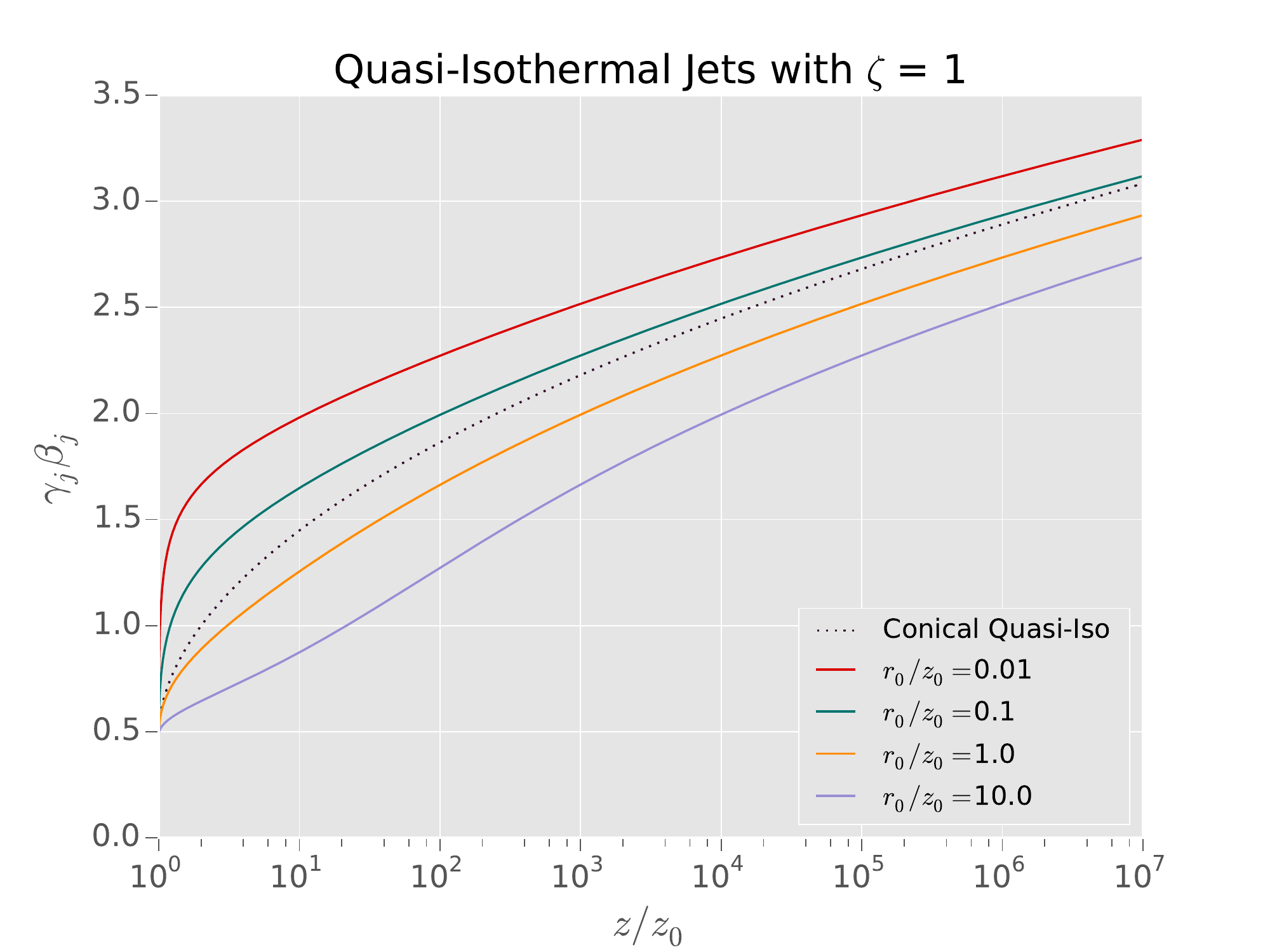}
\caption{This figure shows how a different initial aspect ratio of the jet, $r_0/z_0$, changes the dynamics of a self-collimated quasi-isothermal jet compared to the conical jet considered in Figure \ref{fig:AGNJETvEuler}. As long as $r_0/z_0\sim 1$, the difference between a conical and self-collimated jet is much less than a factor of 2.}
\label{fig:Self-Consistent}
\end{figure}

The assumption of a lateral expansion at constant speed while accelerating vertically results in a jet that is self-collimating. The effect of the self-collimation on the dynamics was not previously considered in \texttt{agnjet}. We show here that the effect is small as long as the initial cross-sectional radius of the jet is roughly equal to the launch height.

The cross-sectional radius of the jet, $r$, is assumed to follow
\begin{equation}
\label{eq:CrossSection}
r = r_0 + (z - z_0)\gamma_0\beta_0 /(\gamma_j\beta_j),
\end{equation}
and conservation of number density of particles is
\begin{equation}
n = n_0 \left(\frac{\gamma_j\beta_j}{\gamma_0\beta_0}\right)^{-1}\left(\frac{r}{r_0}\right)^{-2}.
\end{equation}
For the quasi-isothermal case, i.e., what is currently used in \texttt{agnjet}, the internal energy profile is
\begin{equation}
U_j = \zeta n_0 m_p c^2 \left(\frac{\gamma_j\beta_j}{\gamma_0\beta_0}\right)^{-\Gamma}\left(\frac{r}{r_0}\right)^{-2}.
\end{equation}
($\zeta = 1$ in \texttt{agnjet}).
With the cross-sectional radius assumed in Equation (\ref{eq:CrossSection}), the 1-D Euler equation becomes:
\begin{multline}
\label{eq:FullEulerWithHeatingAndCollimation}
\left\{\gamma_j\beta_j\frac{\Gamma+\xi}{\Gamma-1}-\Gamma \gamma_j\beta_j-\frac{\Gamma}{\gamma_j\beta_j}+\frac{2(z-z_0)\gamma_0\beta_0/(\gamma_j\beta_j)}{r_0\gamma_j\beta_j+\gamma_0\beta_0(z-z_0)}\right\} \\ 
\qquad \times \frac{\partial \gamma_j\beta_j}{\partial z} = \frac{2\gamma_0\beta_0}{r_0\gamma_j\beta_j+\gamma_0\beta_0(z-z_0)};
\end{multline}
and as before,
\begin{equation}
\xi = \frac{1}{\zeta}\left(\frac{\gamma_j\beta_j}{\gamma_0\beta_0}\right)^{\Gamma-1}; \qquad \gamma_0\beta_0 = \sqrt{\frac{\zeta\Gamma(\Gamma-1)}{1+2\zeta\Gamma-\zeta\Gamma^2}}.
\end{equation}
There is now another free parameter in the velocity profile: the initial aspect ratio of the jet: $r_0/z_0$. We show the effects of this new parameter and compare our results to a quasi-isothermal conical jet in Figure \ref{fig:Self-Consistent}. 

%For completeness in the adiabatic case, the 1D Euler equation is:
%\begin{multline}
%\label{eq:FullEulerAdiabaticAndCollimation}
%\left\{\gamma_j\beta_j\frac{\Gamma+\xi}{\Gamma-1}-\Gamma \gamma_j\beta_j-\frac{\Gamma}{\gamma_j\beta_j}+\frac{2\Gamma(1+\gamma^2_j\beta^2_j)(z-z_0)\gamma_0\beta_0}{r_0\gamma^2_j\beta^2_j+\gamma_0\beta_0\gamma_j\beta_j(z-z_0)}\right\} \\
%\qquad \times \frac{\partial \gamma_j\beta_j}{\partial z} = \frac{2\Gamma\gamma_0\beta_0\left(1+\gamma_j^2\beta_j^2\right)}{r_0\gamma_j\beta_j+\gamma_0\beta_0(z-z_0)};
%\end{multline}
%\begin{equation}
%\xi = \frac{1}{\zeta}\left(\frac{\gamma_j\beta_j}{\gamma_0\beta_0}\right)^{\Gamma-1}\left(\frac{z}{z_0}\right)^{2(\Gamma-1)}; \qquad \gamma_0\beta_0 = \sqrt{\frac{\zeta\Gamma(\Gamma-1)}{1+2\zeta\Gamma-\zeta\Gamma^2}}
%\end{equation}

To divvy up the total internal energy into an electron energy density and magnetic energy density, \texttt{agnjet} uses a free parameter $k$, defined as the ratio of the magnetic energy density to the electron energy density. The dependence of the magnetic field on the height is 
\begin{equation}
B = \sqrt{\frac{8 \pi k U_j}{1+k}}.
\end{equation}
$U_j$ has a height dependence that is different if the jet is assumed to be isothermal, quasi-isothermal, or adiabatic. In the isothermal jet, $B\propto r^{-1}(\gamma_j\beta_j)^{-1/2}$, which is slightly slower than the expected dependence if there is flux conservation of a toroidal magnetic field ($\propto r^{-1}(\gamma_j\beta_j)^{-1}$). In the adiabatic or quasi-isothermal case, the magnetic field decreases faster than in an isothermal jet.
The electron's characteristic Lorentz factor $\gamma_{e}$ has the same distance and bulk Lorentz factor dependence as the temperature. $\gamma_{e,0}$ is a free parameter. The density profile of the electrons is determined by number conservation, but the initial number of electrons and positrons is fixed by requiring that $U_{j,0} = n_p m_p c^2$, 
\begin{equation}\label{eq:pair_ratio}
\frac{n_e}{n_p} = \frac{1}{1+k}\frac{m_p}{\gamma_{e,0}m_e}.
\end{equation}
The above relationship will break down if $k$ is too large, and it results in a non-physical, electrostatically charged jet with $n_e<n_p$. $k$ must satisfy the following inequality
\begin{equation}\label{eq:charge_inequality}
k+1\lesssim 110\left(\frac{T_{e,0}}{10^{11}\ \mathrm{K}}\right)^{-1}
\end{equation}
If \texttt{agnjet} requires a $k$ large enough to violate the above inequality to fit the spectrum of an object, the $k$ is inconsistent with the model of \texttt{agnjet}. In this case, it likely means that the jet is Poynting dominated, i.e., $U_B > U_p$, a scenario which is not considered in \texttt{agnjet}.  We note that versions of \texttt{agnjet} prior to 2014 did not use eq (\ref{eq:pair_ratio}) to set the number of electrons and positrons, and instead simply set $n_e = n_p$. In the earlier version, \texttt{agnjet} was only self-consistent if $1+k$ was equal to the right hand side of eq (\ref{eq:charge_inequality}). If we force $n_e=n_p$, as in the earlier version of \texttt{agnjet}, when $1+k <  110\left[T_{e,0}/(10^{11}\ \mathrm{K})\right]^{-1}$,  the difference from a self-consistent solution can be estimated by solving the Euler equation in eq (\ref{eq:FullEulerWithHeatingAndCollimation}) with a $\zeta = (1+k)\left[T_{e,0}/(10^{11}\ \mathrm{K})\right]/110$. The differences are likely to be minor even if $\zeta\ll 1$ because the jet is not accelerated very much even when $\zeta = 1$. If in the previous version of \texttt{agnjet} a solution was found with $1+k \gg 110\left[T_{e,0}/(10^{11}\ \mathrm{K})\right]^{-1}$, the jet is Poynting dominated, and its dynamics are not correctly calculated by \texttt{agnjet}.
\begin{figure}
\includegraphics[width = 0.5\textwidth]{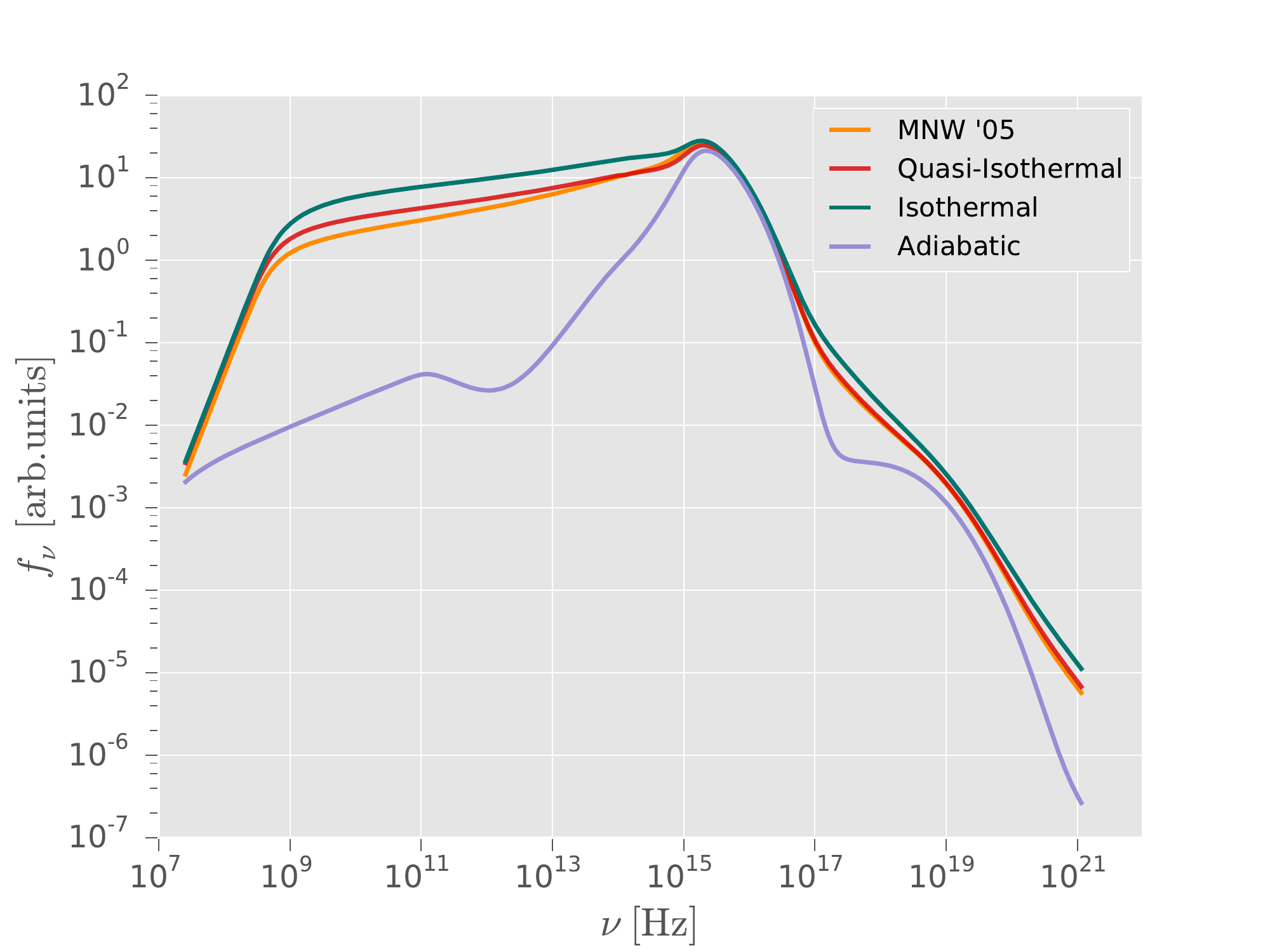}
\caption{This figure shows the different spectral energy distribution (SED) calculated for conical jets in \texttt{agnjet}. The colors correspond to the same Lorentz factor profiles in Figure \ref{fig:AGNJETvEuler}. The models where the temperature is constant or nearly constant all show similar SED, while the adiabatic model has a steep rise below the thermal synchrotron bump.}
\label{fig:SED}
\end{figure}

We plot an example multiwavelength spectral-energy distribution using \texttt{agnjet} in Figure \ref{fig:SED}. In the figure it is clear that the quasi-isothermal, isothermal, and previous \texttt{agnjet} with minor errors all give roughly the same result: a nearly flat, self-absorbed synchrotron spectrum at frequencies below the thermal bump, and a Compton hump in the X-rays. In Figure \ref{fig:SED}, we assume the jet to be self-collimating when calculating the dependence of the internal energy of the jet with height, but we use the velocity profiles from Figure \ref{fig:AGNJETvEuler}. Unlike the models with a constant or nearly-constant temperature throughout the jet, the adiabatic model shows a steep rise from the self-absorbed radio emission to the thermal synchrotron bump. We find that to have a flat radio spectra the jet must be kept at a nearly constant temperature, a similar conclusion as found by \citet{1979ApJ...232...34B} when fitting the radio cores of AGNs or as found by \citet{2013A&A...559L...3M} when fitting the radio spectrum of Sgr A*, or as found by \citet{2009ApJ...699.1919P} when fitting the radio emission of X-ray binaries.

\section{Summary and Discussion}
In this work, we have re-analyzed the hydrodynamical jets derived in \citet{1996ApJ...464L..67F}. When deriving the Lorentz factor profile, \citet{1996ApJ...464L..67F} used the  \textit{maximal jet} model from \citet{1995A&A...293..665F}, that contains an algebraic error. The \textit{maximal jet} assumption has two main conclusions---the jet's power is dominated by its kinetic power, and the jet power must be less than the mass accretion rate of the disk onto the black hole times the bulk Lorentz factor of the jet. We argued in this letter that the second conclusion is likely true in general for astrophysical jets. Even in jets that extract their energy from the black-hole spin, the energy in the jet does not greatly exceed $\dot{M}_{BH} c^2$ because the amount of magnetic flux that can be carried to the black hole depends on the mass accretion rate. The  maximal efficiency of jet production found in GRMHD simulations is $\lesssim 300$\% \citep{2011MNRAS.418L..79T, 2015MNRAS.449..316N}. Furthermore, while it is true that the jet's power need not be dominated by its kinetic power, we argued that this is a good approximation in a jet with a sufficiently small opening angle and a small terminal bulk Lorentz factor. X-ray binaries and low-luminosity AGN likely host such jets, so models based on these assumptions are well-founded for such objects. In addition, we corrected minor algebraic mistakes made in the derivation of the Lorentz factor profile from \citet{1996ApJ...464L..67F}. The effects of correcting these errors are to make the quasi-isothermal jet behave more similarly to a jet where the temperature is kept completely constant.

The Lorentz factor profile from \citet{1996ApJ...464L..67F} was used in the \texttt{agnjet} model described in \citet{2005ApJ...635.1203M} \& \citet{2009MNRAS.398.1638M}. \texttt{agnjet} has been used to fit the multiwavelength spectra of outflow dominated x-ray binaries and nearby low-luminosity active galactic nuclei. In \texttt{agnjet}, the jet is assumed to be self-collimating, but the collimation was not self-consistently applied to the jet's dynamics. We show that the effects of the self-collimation are negligible for quasi-isothermal jets as long as the aspect ratio  of the jet at the launching point is of order unity. We examined the effects of the new Lorentz factor profile on the spectral energy distribution calculated by \texttt{agnjet}. We find there is very little difference between the assumed quasi-isothermal jet and a completely isothermal jet, however, we find a large difference in the spectral energy distribution between an adiabatic jet and a jet where the temperature is kept roughly constant. We find that isothermal jets are required to match the flat radio spectra seen in hard/quiescent state XRB and low-luminosity AGN. We do not self-consistently account for how the jet is kept hot. If the jet is heated internally, the heating mechanism would change the  Lorentz factor profile from the one calculated in this work. If the gas is shock heated, the shocks will change the jet's momentum, and if the jet is Poynting dominated, the dissipation of the magnetic field to heat the particles will change the magnetic pressure gradient.

Finally, we note that in addition to this work, there is an ongoing effort to derive a MHD-consistent jet model that will be capable of calculating the jet properties self-consistently, e.g. \citet{2010ApJ...723.1343P, 2013MNRAS.428..587P, 2014MNRAS.438..959P}, Ceccobello et al. \textit{in prep.} Such models will be able to address many of the short-comings of the models described in this work.

\begin{acknowledgements}
PC would like to thank Heino Falcke, Peter Biermann,  Sera Markoff, Pawan Kumar, Rodolfo Barniol Duran, and Thomas D. Russell for valuable discussions. PC acknowledges financial support from the WARP program of the Netherlands Organisation for Scientific Research (NWO). CC and RC acknowledge NWO/Nova. YC is suported by the European Union’s Horizon 2020 research and innovation programme under the Marie Sklodowska-Curie Global Fellowship grant agreement No 703916.
\end{acknowledgements}
\bibliographystyle{aa}
\bibliography{Euler.bib}

\end{document}